\documentclass[oldversion]{aa}

\usepackage{epsfig}
\usepackage{graphics}
\usepackage{float}
\usepackage{amsmath}
\usepackage{multirow}
\usepackage{longtable}
\usepackage{rotate}
\usepackage{array}
\usepackage{subfigure}
\def\kms{\ifmmode{\rm km\,s^{-1}}\else\hbox{$\rm km\,s^{-1}$}\fi}

\setlongtables

\begin{document}

\title{Bayesian inference for orbital eccentricities}

\author{L.B.Lucy}
\offprints{L.B.Lucy}
\institute{Astrophysics Group, Blackett Laboratory, Imperial College 
London, Prince Consort Road, London SW7 2AZ, UK}
\date{Received ; Accepted }

\abstract{Highest posterior density intervals (HPDI's)
are derived for the true eccentricities $\varepsilon$ of spectroscopic 
binaries with measured values $e \approx 0$. These 
yield upper limits when $e$ is below the detection threshold $e_{th}$  
and seamlessly transform to upper and lower bounds when $e > e_{th}$.  
In the main text, HPDI's are computed with 
an {\em informative} eccentricity prior representing orbital decay due to tidal 
dissipation. 
In an appendix, the corresponding HPDI's are computed with  
a uniform prior and are the basis for a   
revised version of the Lucy-Sweeney test, with the previous outcome 
$\varepsilon = 0$ now replaced by an upper limit $\varepsilon_{U}$. 
Sampling experiments with known prior confirm the validity of the HPDI's.
}

\authorrunning{Lucy}
\titlerunning{Orbital eccentricities}
\maketitle

\section{Introduction}

For over two centuries, astronomers have been able to detect and analyse
orbital motions for objects beyond the solar system.
From measured positions (visual binaries) or
radial velocities (spectroscopic binaries, exoplanets), 
orbital elements and their standard errors are typically obtained by 
least-squares.
Accordingly, at this late date, 
when orbital motion is detected, {\em prior} information concerning
similar objects is available and can be incorporated into the analysis. 

An example is the Lucy-Sweeney (1971; LS)
test for the statistical significance of small measured eccentricities $e$
for spectroscopic binaries (SB's). 
The prior information that provided support for the LS test was as follows:\\   
1) The small $e$'s (typically $\la 0.05$) of numerous catalogued 
SB's were $\sim E(e|0)$, the expected
value of $e$ due to measurement errors when the true value is 
$\varepsilon = 0$.\\
2) Savedoff's (1951) investigation of $e \: cos \: \omega$, where 
$\omega$ is the longitude of periastron. If an SB is also an eclipsing binary
(EB), $e \: cos \: \omega$ can be determined from the velocity curve
($sp$) and 
{\em independently} from the light curve ($ph$).
For EB's with 
secondary eclipses midway between consecutive primary eclipses,
$(e \: cos \: \omega)_{ph} = 0$, but the values of $(e \: cos \: \omega)_{sp}$
are scattered over the interval (-0.05, 0.05), confirming that non-zero
$e$'s $\la 0.05$ are often spurious.\\    
3) Tidal dissipation gives $\varepsilon \rightarrow 0$ as 
$ t \rightarrow \infty$ since, for two point masses,  $\varepsilon = 0$ is 
the state of minimum orbital
energy for fixed orbital angular momentum. Thus, if the time constant 
for this decay is short enough, the system is likely to be observed
when $\varepsilon \ll E(e|0)$ - i.e., well below the measurement 
threshold.   

In view of this prior information, LS adopted $\varepsilon = 0$ as the 
preferred (null) hypothesis ($H_{0}$) and imposed a moderately demanding 
level of significance before rejecting $H_{0}$ and accepting
an elliptical orbit.

Given that secular evolution due to tidal dissipation
was already well established in 1971 and is not less so now, there is
merit in explicitly incorporating this mechanism into the analysis
rather than implicitly via the LS preference for 
$\varepsilon = 0$. This can be achieved by replacing the
{\em frequentist} approach of LS by one based on Bayes' theorem.

\section{Estimating the true eccentricity $\varepsilon$}

We suppose 
that radial velocities of a single-lined
spectoscopic binary (SB1) have been analysed to estimate the orbital
elements and their standard errors. We ask: what can be inferred about the 
error-free elements when orbital evolution due to
tidal dissipation is taken into account? 

\subsection{Posterior probability} 

We adopt the notation used in Lucy(1974; L74). The vectors of 
the estimated
and the error-free elements are denoted by $\vec{x}$ and $\vec{\xi}$, 
respectively; and the
distribution of probability
in $\vec{x}$-space for given $\vec{\xi}$ is denoted by
$\Pi(\vec{x}|\vec{\xi}) d\vec{x}$. Integrating over $\vec{\xi}$-space, we 
find that  
the distribution of probability in $\vec{x}$-space is $\phi(\vec{x}) d\vec{x}$,
where
\begin{equation}
  \phi(\vec{x}) = \int \psi(\vec{\xi}) \Pi(\vec{x}|\vec{\xi}) d \vec{\xi}
\end{equation}
Here $\psi$ is the probability density function (pdf) that 
represents our 
{\em prior} knowledge about $\psi(\vec{\xi}) d \vec{\xi}$, the
distribution of probability in $\vec{\xi}$-space for the SB1's true elements.

Since SB1's with small $e$'s are of interest, we suppose that 
Sterne's (1941) elements have been chosen, as in LS. The six elements are 
then: $P$, 
the orbital period; $\gamma$, the systemic velocity; $K$, the semi-amplitude
of the velocity curve;
$T_{0}$, an epoch at which the mean longitude is zero; and the pair
$e \: cos \: \omega, \;  e \: sin \: \omega$. In terms of these parameters,
the radial velocity curve of an SB1 is given by Eq.(1) in LS.

With this choice and the assumption $e^{2} \ll 1$, the off-diagonal elements
of the least-squares matrix
have zero expectation values when observational weight is uniformly
distributed in phase (LS). Since observers strive to meet this
condition, we assume it to be true; and this then implies negligible
correlations between the elements. Accordingly, to a good approximation,
the error-broadening kernel $\Pi(\vec{x}|\vec{\xi})$ is simply the product of
the six gaussians giving the independent error distributions of the six 
elements. 
 
Next consider the pdf $\psi (\vec{\xi})$, which is convolved in Eq.(1) with the
kernel $\Pi(\vec{x}|\vec{\xi})$. It follows that $\psi$ is only relevant if it
varies significantly within the error bars of an individual orbital element.
This is {\em not} true for $P, \gamma, K$, and $T_{0}$. 
However, when orbit circularisation is taken into account, $\psi(\vec{\xi})$ 
may vary significantly within the domain 
$(e \: cos \: \omega \pm \sigma_{e cos \omega},
 e \: sin \: \omega \pm \sigma_{e sin \omega})$.    
Accordingly, we now integrate Eq.(1) with respect to the
($P,\gamma,K,T_{0}$)-components of $\vec{x}$ and use the normalization
of $\Pi(\vec{x}|\vec{\xi})d\vec{x}$ with respect to each of these four elements.
The result is 
\begin{equation}
  \phi_{c}(x,y) = \int \psi_{c}(\xi,\eta) \Pi_{c}(x-\xi,y-\eta) d \xi d \eta
\end{equation}
where $x = e \: cos \: \omega$ and $y = e \: sin \: \omega$ are the two 
surviving estimated
quantities, whose error-free values are $\xi = \varepsilon \: cos \: \varpi$
and $\eta = \varepsilon \: sin \: \varpi$, respectively, and the subscript
$c$ indicates that the pdf's are defined on a Cartesian grid.      

We now assume that the standard errors of $e \: cos \: \omega$ and  
$e \: sin \: \omega$
are equal, which
is true of their expectation values when observational
weight is uniformly distributed in phase (LS). The broadening kernel is then the
circular normal distribution 
\begin{equation}
  \Pi_{c}(x-\xi,y-\eta) = \frac{1}{2 \pi \mu^{2}}
   exp \left[-\frac{(x-\xi)^{2}+(y-\eta)^{2}}{2 \mu^{2}} \right]
\end{equation}
where $\mu = \sigma_{e cos \omega} = \sigma_{e sin \omega}$.

Bayes' theorem can now be invoked to derive - see Eq.(10) in L74 -  the
posterior probability 
\begin{equation}
  Q_{c}(\xi,\eta|x,y) = \frac{\psi_{c}(\xi,\eta) 
                               \Pi_{c}(x-\xi,y-\eta)}{\phi_{c}(x,y)}
\end{equation}
Thus, if the least-square solution is
\begin{equation}
   x = e \: cos \: \omega \pm \mu, \;\;\;
                                     y = e \: sin \: \omega \pm \mu
\end{equation}
the distribution of probability in 
($\varepsilon \: cos \: \varpi, \varepsilon \: sin \: \varpi$)-space is
given by  $Q_{c}(\xi,\eta|x,y)d\xi d\eta$. 
To evaluate this {\em posterior} distribution, we must specify what our
expectations were for $\psi_{c}(\xi,\eta) d \xi d \eta$,
the {\em prior} distribution of probability in 
($\varepsilon \: cos \: \varpi, \varepsilon \: sin \: \varpi$)-space.

For the problem at hand, polar ($p$) coordinates ($\varepsilon, \varpi$)
are more convenient than the Cartesian coordinates
$(\xi, \eta)$ in the above formulae. The Jacobian
of the transformation is $J = \varepsilon$, so that, by conservation of
probability, 
\begin{equation}
          \psi_{p}(\varepsilon, \varpi) =\varepsilon \: \psi_{c}(\xi, \eta)
\end{equation}
and 
\begin{equation}
          Q_{p}(\varepsilon,\varpi|x,y) = \varepsilon \: Q_{c}(\xi,\eta|x,y)
\end{equation}

\subsection{A physical model for $\psi_{p}$} 

Given $P$, there is a 2-D family of binaries that match 
the measured $K$ at some inclination. For simplicity, a representative example
is chosen to avoid integrating over all possibilities. 

Now consider an ensemble of such binaries 
that form at a uniform rate in the solar
neighbourhood and have lifetime $t_{*}$. We further suppose that all have 
$\varepsilon = \varepsilon_{0}$ at $t = 0$ and that thereafter $\varepsilon$
decays exponentially with fixed $e$-folding time $t_{*}/\nu$, so that
\begin{equation}
  \ln \: \varepsilon (t)    =  \ln \: \varepsilon_{0} \: - \: 
                         \nu \: \left(\frac{t}{t_{*}}\right)  
\end{equation}
It follows that $\ln \varepsilon$ is uniformly distributed in
the interval ($\ln \varepsilon_{*}, \ln \varepsilon_{0}$), where
$\varepsilon_{*} = \varepsilon_{0} exp(-\nu)$. 
The probability that 
$\varepsilon \in (\varepsilon,\varepsilon + d \varepsilon)$ is therefore  
$d \varepsilon/\nu \varepsilon$.  

If we now assume randomly oriented orbits, the probability
that $\varpi \in (\varpi,\varpi + d\varpi)$ is $d\varpi/2 \pi $. Accordingly,
the prior probability that a binary is in the element $d\varepsilon d\varpi$ 
at $(\varepsilon,\varpi)$ is $\psi_{p} d\varepsilon d\varpi$, where
\begin{equation}
  \psi_{p}(\varepsilon,\varpi) = \frac{1}{2 \pi \nu \varepsilon} 
     \;\; for \;\;  \varepsilon \in (\varepsilon_{*},\varepsilon_{0})
\end{equation}
Note that $\psi_{p}$ decreases
with increasing $\nu = \ln (\varepsilon_{0}/\varepsilon_{*})$. Nevertheless, 
normalization of this pdf is maintained 
by the corresponding decrease in $\varepsilon_{*}$, the lower limit for
integrations over $\varepsilon$.

In introducing this physical model, we in effect adopt an {\em informative} 
prior. The following quote is apt: " The real power of Bayesian inference 
lies in its ability to incorporate 'informative' prior information, not
'ignorance' " (Feldman \& Cousins 1998).

\subsection{Distribition of $\varepsilon$} 

For numerical calculations, it is convenient to transform the integral
in Eq.(2) into an integration with respect to the polar coordinates
$\varepsilon$ and $\varpi$, so that
\begin{equation}
  \phi_{c}(x,y) = \int_{\varepsilon_{*}}^{\varepsilon_{0}} \int_{0}^{2 \pi}
      \psi_{p}(\varepsilon,\varpi) \Pi_{c}(x-\xi,y-\eta)
      \: d \varepsilon d \varpi
\end{equation}
and to use logarithmic spacing in $\varepsilon$ in order to accurately 
evaluate the contribution near $\varepsilon_{*}$. Note that $\phi_{c}$ is
independent of $\omega$ when $\psi_{p}$ is independent of $\varpi$, 
as in Eq.(9).

With $\phi_{c}$ evaluated, the distribution of probability in
($\varepsilon,\varpi)$-space is given by 
$Q_{p}(\varepsilon,\varpi|x,y) d \varepsilon d \varpi$, with $Q_{p} = \varepsilon Q_{c}$ 
from Eq.(4). This 2-D pdf, which in general is {\em not} independent of 
$\varpi$, may be of interest when analysing a particular SB1. But here our 
interest is in $\varepsilon$, so we integrate over $\varpi$ to obtain  
\begin{equation}
  q(\varepsilon|e) = \int_{0}^{2 \pi} Q_{p}(\varepsilon,\varpi|x,y) \: d \varpi
 \end{equation}
Note that $q(\varepsilon|e)$ is independent of $\omega$ because of the absence
of a correlation term in $\Pi_{c}$ - see Eq.(3). This in turn follows from 
the assumptions (Sect.2.1) that $e^{2} \ll 1$ {\em and} that observational
weight is uniformly distributed in phase.
    
The posterior probability that the true eccentricity
$ \in (\varepsilon, \varepsilon+ d \varepsilon)$ is therefore   
$q(\varepsilon|e) d \varepsilon$, with mean value
\begin{equation}
   <\varepsilon>  \: = \: \int_{\varepsilon_{*}}^{\varepsilon_{0}}
                           \varepsilon \: q(\varepsilon|e) d\varepsilon
\end{equation} 

\subsection{Bayesian terminology}

In the above, notation and terminology is from L74. To modern Bayesians, 
$\Pi_{c}(x-\xi,y-\eta)$ is the likelihood and 
$\phi_{c}(x,y)$ is the Bayes' factor. Elsewhere, modern usage is
followed with respect to the 
terms prior pdf, posterior pdf and credible intervals.

\section{Numerical results}

The theory of Sect.2 is now illustrated by computing a
particular case in detail.

\subsection{Parameters}

There are two basic parameters, $e/\mu$ and $\nu$, the number of e-folding 
decay times in $t_{*}$.

Note that $\epsilon_{0}$ is not a consequential parameter provided that
$e \ll \epsilon_{0}$. In effect, we assume
that an SB1 with $e \approx 0$ has reached this
configuration due to secular evolution and not due to the
formation mechanism. In these calculations,  $\epsilon_{0} = 0.5$.

We choose $\nu = 8.52$, so that $\varepsilon_{*} = 10^{-4}$. Then, with
$\mu = 0.01$, $\varepsilon(t) < 2.45 \mu$,
the LS threshold, when $t > 0.35 t_{*}$. Thus, from our ensemble of
SB1's, $\approx 65\%$ would be assigned $\varepsilon = 0$ by the LS test.  

\subsection{The posterior pdf $ \;  \chi(\log \varepsilon|e)$}

Because of the concentration of probability towards $\varepsilon_{*}$, plots
are more informative if the abscissa is $\log \varepsilon$ rather than
$\varepsilon$. Accordingly, we define
\begin{equation}
  \chi(\log \varepsilon|e) \: = \: \varepsilon \: q(\varepsilon|e) 
                                                         \times \ln 10   
\end{equation}
In Fig.1, this pdf is plotted for $e/\mu = 1.0, 2.45, 3.03$ and $3.72$,
values selected as follows:
If $\varepsilon = 0$, LS showed that the probability $p_{e}$
of exceeding $e$ is given by 
\begin{equation}
   \ln \: p_{e} = - \frac{1}{2} \: \left( \frac{e}{\mu} \right)^{2} 
\end{equation}
provided that $\mu \ll 1$.  
Therefore, when testing $H_{0}$, the above values of $e/\mu$
correspond to levels of significance 61, 5, 1 and 0.1 $\%$,
respectively. The criterion $e/\mu > 1$ was proposed and implemented 
by Luyten (1936); the 5$\%$ level by LS. 

For $e/\mu = 3.72$, $\chi$ is an asymmetric
bell-shaped function peaking at $\approx e$, but with a 
tail 
extending down to $\log \varepsilon_{*} = -4.0$. As $e/\mu$ decreases, 
the peak weakens and the tail
strengthens. At $e/\mu = 1.0$, the peak is absent and all
the probability is in the tail, which derives from the physical model.   
Intermediate calculations show that the peak first appears at
$e/\mu = 1.42$. Thus, for $e/\mu < 1.42$, $\chi$ is a monotonically 
decreasing
function of $\varepsilon$. For $e/\mu > 1.42$, $\chi$ is unimodal.

The pdf $\chi$ for $e/\mu = 3.72$ and $\mu = 0.01$ in Fig.1 is computed 
with $\varepsilon_{0}/\mu = 50$. Repeating this calculation shows that
$\chi$ is independent of the upper limit $\varepsilon_{0}$ provided that
$\varepsilon_{0}/\mu \ga 10$.

\begin{figure}
\vspace{8.2cm}
\includegraphics{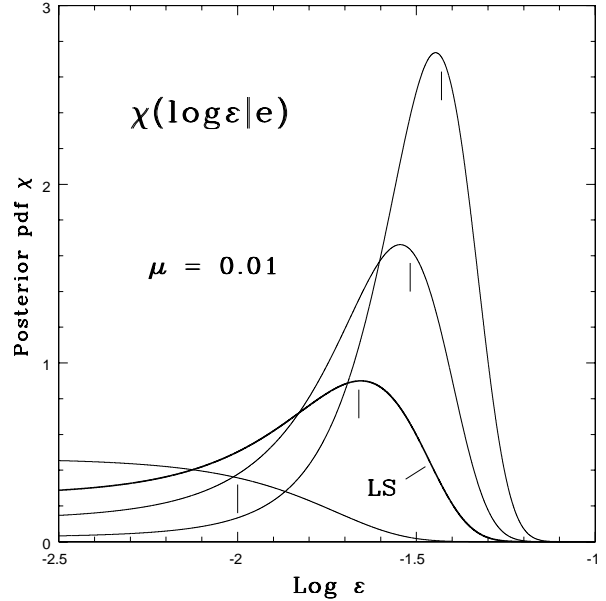}
\caption{The pdf $\chi(\log \varepsilon|e)$ for the values of $e$ indicated
by the vertical lines. These are at $e/\mu = 1.00, 2.45, 3.03$ and
$3.72$, corresponding to levels of significance of 61, 5, 1, and 0.1$\%$,
respectively. The bold curve is the pdf at the point where the LS-test 
switches from accepting to rejecting a circular orbit.}  
\end{figure}

\subsection{Percentiles}

The posterior pdf's in Fig.1 imply asymmetric and rapidly changing credible 
(or Bayesian confidence) intervals as $e/ \mu$ varies. These are plotted in
Fig.2 for the indicated
values of the probability that the true eccentricity is 
$ < \varepsilon$. For the normal distribution, the values 
$0.159, 0.500, 0.841$ and $0.977$
correspond to displacements of $-1, 0, +1$ and $+2\sigma$, respectively.

\begin{figure}
\vspace{8.2cm}
\includegraphics{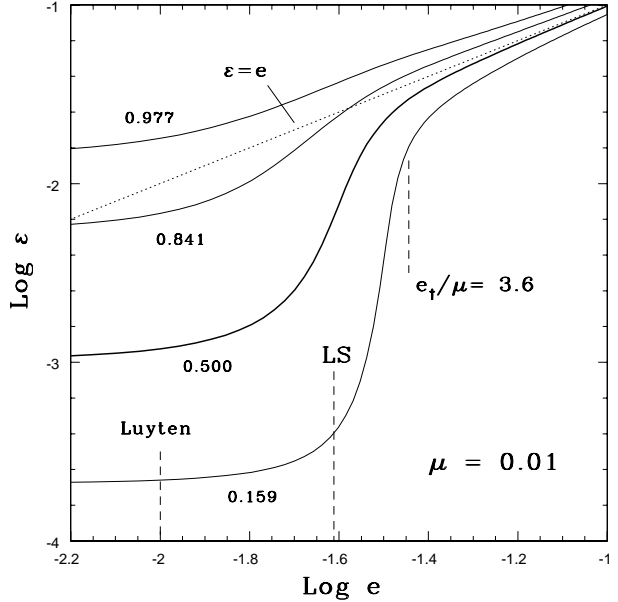}
\caption{Credible intervals for the posterior pdf  $\chi(\log \varepsilon|e)$ 
as functions of $e$. The plotted boundaries 
correspond to the true eccentricity being 
$ < \varepsilon$ with the indicated probabilities. 
The values of $e$ corresponding to
levels of significance 61$\%$ (Luyten) and 5$\%$ (LS) 
are indicated, as is the regime transition at $e_{\dagger}$.}
\end{figure}

Fig.2 reveals a dramatic switch in solution regime at 
$e_{\dagger}/\mu \approx 3.6$. For
$e \ga e_{\dagger}$, the measured value $e$ is close to the $50\%$ percentile
and is tightly enclosed by the $\pm 1\sigma$ intervals. In this regime,
inferences are dominated by the actual measurement $e \pm \mu$. 
But this ceases to be so for $e \la e_{\dagger}$.
Thus, for $e/\mu < 2.65$,
the 'solution' $\varepsilon = e$ falls outside the $\pm 1\sigma$ intervals.
Evidently, for $e \la e_{\dagger}$, 
inferences are increasingly dominated by the model of Sect.2.2.

\subsection{Highest posterior density intervals}

For $e/\mu < 1.42$, $\chi$ is a monotonically decreasing function
of $\varepsilon$ (Sect. 3.2). It follows that traditional, {\em equal-tail}
credible intervals 
{\em exclude} the point ($\varepsilon = \varepsilon_{*}$) with greatest 
probability density (pd).
This undesirable feature is avoided by instead computing 
{\em highest posterior density intervals} (HPDI; Box \& Tiao 1973).  
These intervals are such that every point included has a higher pd than
every point excluded.

In general, the calculation of HPDI's is non-trivial. 
But here the pdf's $\chi$ are not pathological (Fig. 1), and so the 
following 
clipping algorithm finds the HPDI for specified $e$
and designated enclosed probability $(1-\alpha)$:\\

Let $\log \varepsilon_{k_{L}}, \dots, \log \varepsilon_{k_{U}}$ 
be consecutive grid
points that belong to and define the HPDI 
($\varepsilon_{k_{L}},\varepsilon_{k_{U}}$).
Then an HPDI with smaller
included probability is obtained by eliminating the grid point
$\log \varepsilon_{k_{L}}$ if $\chi_{k_{L}} < \chi_{k_{U}}$ or the grid point
$\log \varepsilon_{k_{U}}$ if $\chi_{k_{U}} < \chi_{k_{L}}$.
This is repeated until the included probability $ = (1-\alpha)$.\\ 
 
The $95\%$ HPDI's thus obtained for  
$\chi(\log \varepsilon|e)$
are plotted for $e = 0.000 (0.001) 0.080$ in Fig.3. For $e < 0.018$, the
HPDI's are effectively one-tail intervals since the lower
bound is $\varepsilon_{L} = \varepsilon_{*} = 10^{-4}$. Accordingly, 
for this problem, HPDI's
provide a seamless transition from upper limits for non-detected 
to two-sided intervals for detected eccentricities
(cf. Feldman \& Cousins 1998). This is an appealing
aspect of HDPI's for interpreting measured $e$'s and their uncertainties.

\begin{figure}
\vspace{8.2cm}
\includegraphics{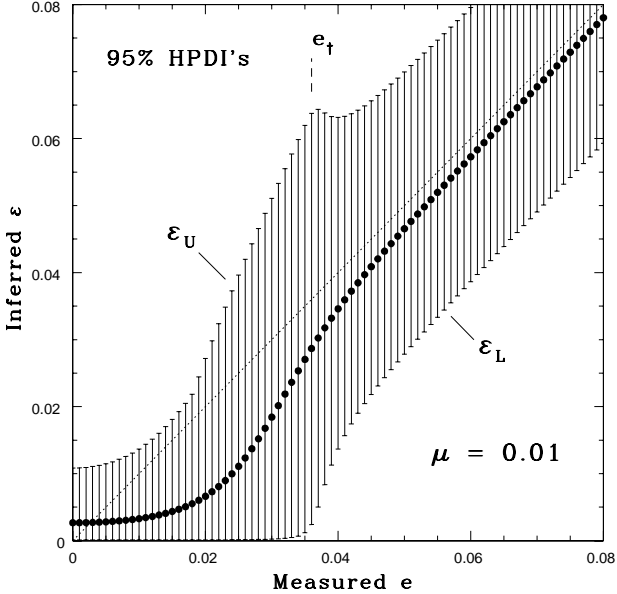}
\caption{Highest posterior density intervals for the posterior pdf
$\chi(\log \varepsilon | e)$. 
The enclosed posterior probability of each HPDI is 
$95\%$, and the prior pdf $\psi$ is given by Eq.(7). The filled circles are 
the posterior means $<\varepsilon>$ computed from Eq.(10).}
\end{figure}

For $e < 0.0142$, the interval excluded from an HPDI is a single-tail
because of the aforementioned monotonicity. For $e > 0.0142$, the pdf $\chi$ is 
unimodal (Fig.1),
with a maximum whose location $\rightarrow e$ as $e/\mu$ increases.
This emerging, {\em measurement-driven} maximum 
eventually brings about the transition from one- to
two-tailed intervals.  
For $e > 0.018$, the excluded probability is contained in two tails,
with the upper tail's probability being initially $5\%$, but this 
decreases to $\approx 2.5\%$ when $e/\mu \gg 1$
because of $\chi$'s increasing symmetry - see Fig.1.

Note that for the normal distribution, ${\cal N}(0,1)$, the $95\%$ 
HPDI is the familiar equal tail interval $(-1.96,+1.96)$ and the
width of this interval is the narrowest that ecloses $95\%$ 
of the probability. For non-symmetric pdf's, HPDI's are the narrowest intervals
enclosing probability $(1-\alpha)$ and as such are a natural generalization
of the conventional equal-tail intervals for symmetric, bell-shaped pdf's. 

Confidence intervals are an economical means of conveying the compactness or
otherwise of a variate's distribution. The resulting loss of information, 
if of concern, 
can be avoided by plotting the pdf's, as in Figs.1 and A.2.

\subsection{Detection threshold}

The transition from one- to two-tailed HPDI's is a natural
definition of the detection threshold $e_{th}$ for non-zero eccentricity - 
i.e., the measured value $e$ above which attribution purely to measurement
errors is
implausible. However, Fig.3 shows that $\varepsilon_{L}$ remains $\ll e$
for a considerable interval beyond $e_{th} = 0.018$, which in any case
depends on $\nu$, a parameter likely to be only crudely estimated. 

If detection is crucial for a subsequent investigation
- e.g., an observing program - then a threshold closer to $e_{\dag}$ should
be adopted (Figs. 2 \& 3).

\subsection{Simulation}

The role that HPDI's can play in reporting eccentricities is best illustrated
by sampling experiments. Accordingly, synthetic data for the model of 
Sect.2.2 are created as follows:\\
If $z_{1},z_{2}$ are random numbers in $(0,1)$, a random ensemble
member in $(\varepsilon,\varpi)$-space is at   
\begin{equation}
   \varepsilon = \varepsilon_{0} \: exp(-\nu z_{1})  \; \;\;\;\;\;
                                  \varpi = 2 \pi \: z_{2} 
\end{equation}
Then, if $\zeta_{1},\zeta_{2}$ are random gaussian variates, this ensemble
member is observed at the point 
\begin{equation}
   e \: cos \: \omega= \varepsilon \: cos \: \varpi + \mu \zeta_{1} \;\;\;\;
   e \: sin \: \omega= \varepsilon \: sin \: \varpi + \mu \zeta_{2}
\end{equation}
Repeated $N$ times, the resulting $e$'s
comprise a simulated observing campaign of $N$ random ensemble members 
whose exact eccentricities $\varepsilon$ are known. 

With $\varepsilon_{*} = 10^{-4}$ , a sample of $N = 1000$ SB1's 
are plotted in Fig. 4. As expected, the large majority of
the points fall within the $95\%$ HPDI's  
($\varepsilon_{L},\varepsilon_{U}$).
With Luyten's criterion $e > \mu$, $79.5\%$ of this sample would have
their elliptical orbits accepted. But Fig.4 clearly shows that most systems
with $e \in (0.01,0.03)$ have $\varepsilon < 0.01$ and so exceed Luyten's
criterion because of the bias of the
non-negative estimator $e$ (LS). With the LS criterion $e > 2.45 \mu$,
the accepted percentage drops to $40.5$, and most of the systems with 
$e \in (0.01,0.03)$ would now be assigned $\varepsilon = 0$, a marked 
improvement.

The further improvement provided by the HPDI's is that the
assignment $\varepsilon = 0$ can now be replaced by an upper limit. Thus, 
for example, with this choice of prior, an SB1 with $e = 0.01 \pm 0.01$  
is preferably reported as $\varepsilon < \varepsilon_{95} =  0.014$.
From the standpoint of testing
theories of tidal dissipation,
an upper limit is more informative than $\varepsilon = 0$.   

The sampling procedure can also be used to validate the
upper limits. A sample with $N = 10^{6}$ has
161,521 systems with $\log e \in (-2.1,-1.9)$ and 8,317 of these have
$\varepsilon > \varepsilon_{95}$. Thus $94.85\%$ lie below the upper limit,
closely agreeing with the designated $95\%$.

\begin{figure}
\vspace{8.2cm}
\includegraphics{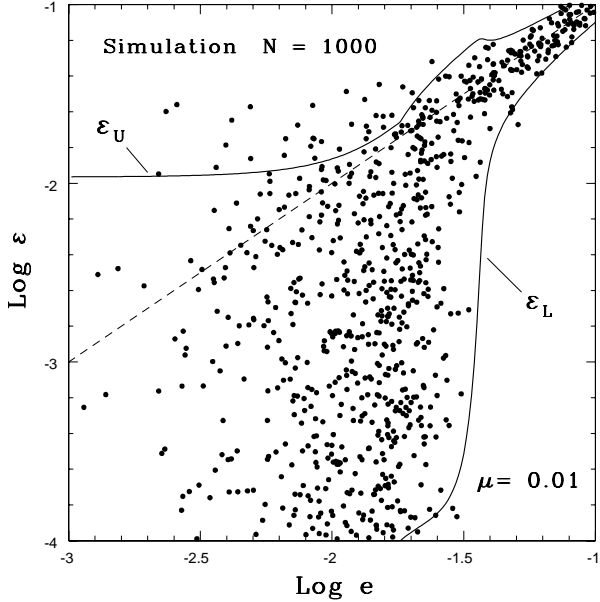}
\caption{Ensemble simulation comprising $N = 1000$ SB1's with the
parameters of Sect.3.1. The plotted points
randomly sample the time interval $(0,t_{*})$, have true eccentricities
$\varepsilon$ from Eq.(6), and have gaussian measurement errors added to derive 
$e$. The $95\%$ HPDI's 
($\varepsilon_{L},\varepsilon_{U}$) are indicated. 
For $e <  e_{th} = 0.018$, the lower
limit $\varepsilon_{L} = \varepsilon_{*} = 10^{-4}$.}  
\end{figure}

\subsection{Upper limits}

Upper limits $\varepsilon_{U}$ when $e < e_{th}$ are the most useful
products of this Bayesian machinery. With their validity confirmed above,
their dependence on the orbital decay rate is now explored.

Consider the representative measurement 
$e = 0.01 \pm 0.01$, so that $e/\mu = 1.0$, well below the detection threshold.
In Table 1, the HPDI $95\%$ upper limits $\varepsilon_{95}$ are given for this 
$e/\mu$ as a function of $\varepsilon_{*}$. 
The decrease of $\varepsilon_{95}$ with increasing $\nu$ reflects statistical 
reality: if the $e$-folding time $t_{*}/\nu \ll t_{*}$, most systems will have 
$\varepsilon \ll e$ and a correspondingly small $\varepsilon_{95}$. Note that
$\varepsilon_{95}$ even drops below $e$ when $\nu \ga 13$.

In analysing an observed system without a good estimate of $\nu$ or,
equivalently, of
$\varepsilon_{*}$, the {\em conservative} approach is to suppose that
$\varepsilon_{*}$ is no more than a factor $\sim 10$ below the measured $e$,
thus avoiding claiming too low an upper limit $\varepsilon_{U}$. An even
more conservative approach is to adopt a uniform prior - see Appendix.

\begin{table}

\caption{HPDI $95\%$ upper limits $\varepsilon_{95}$ when $e/\mu = 1.0$.}

\label{table:1}

\centering

\begin{tabular}{c c c }

\hline\hline

 $\log \varepsilon_{*} $   & $ \nu$   & $ \varepsilon_{95}/\mu$  \\

\hline
\hline

    -3.0     &    6.21   &   1.71  \\

    -4.0     &    8.52   &   1.37  \\

    -5.0     &   10.82   &   1.15  \\

    -6.0     &   13.12   &   0.98  \\

    -7.0     &   15.42   &   0.85  \\

    -8.0     &   17.73   &   0.74  \\

    -9.0     &   20.03   &   0.65  \\

\hline
\hline

\end{tabular}

\end{table}

\section{Conclusion}

In this paper, by incorporating a model of an SB1's secular evolution,
Bayes' theorem is used to infer bounds on its
exact eccentricity $\varepsilon$ given its measured value $e \pm \mu$.
Because the system's lifetime $t_{*}$ is finite, the asymptote
$\varepsilon = 0$ is never reached. Thus, in contrast to Luyten(1936)
and LS, the statistical problem is not one of model selection. Systems
assigned $\varepsilon = 0$ by these earlier tests should
preferably have upper limits $\varepsilon_{U}$ computed. 

As Fig.4 shows, the Bayesian upper limits contain the systems for which $e$ is
significantly larger than $\varepsilon$ due to measurement errors and bias.
Thus, a major historical cause of spurious $e$'s is eliminated.
But physical causes remain, 
such as those due to proximity effects or to additional line absorption by 
gas streams.
An example is $\zeta$ TrA, for which Skuljan et al. (2004)
improved the
precision of the radial velocities by a remarkable factor of 100 and
reported a small but highly significant $e = 0.0140 \pm 0.0002$.
However, the significant non-detection of the Keplerian third harmonic
(Lucy 2005) invalidated this claim\footnote{Hearnshaw et al. (2012) have
just reported {\em eleven} additional non-detections.}. As precision improves,
similar testing
for the 
third and higher Keplerian harmonics is essential for 
confirming that an orbit is truly eccentric. 
In addition, an update of Savedoff's (1951) work
would provide numerous examples of spurious $e$'s for investigation
into physical causes other than measurement bias.

\appendix

\section{Uniform prior}

The model of Sect.2.2 is {\em not} appropriate if the SB or
star-planet system has additional components causing significant 
gravitational perturbations. In this circumstance, a sensible option is to 
assume a uniform prior for $\varepsilon$,
as is already common practice for exoplanets (e.g., Ford 2006; 
Eastman et al. 2012). Together with the assumption of randomly oriented 
orbits,  the prior probability of the system being in 
$d \varepsilon d \varpi $ is then $\psi_{p} d \varepsilon d \varpi $,
where
\begin{equation}
  \psi_{p}(\varepsilon, \varpi)  =  \frac{1}{2 \pi} 
\end{equation} 
which now replaces Eq.(9) in Sects. 2.3 \& 3.

From the resulting posterior pdf $q(\varepsilon|e)$, the $95\%$ HPDI's and
means $<\varepsilon>$ are plotted in Fig.A.1 for $e \in (0.00, 0.08)$. 
Comparison with Fig.3 shows that the HPDI's are nearly identical for 
$e/\mu \ga 5$. However, for $e/\mu \la 2$ - i.e., in the non-detection 
domain - the upper limits $\varepsilon_{U}$ in Fig.3 are markedly lower, 
reflecting the effect of tidal circularisation in creating systems
with $\varepsilon \ll e$.   

In contrast with Fig.3, Fig.A.1 shows that a uniform prior results in a
sharply defined detection threshold $e_{th}/\mu  = 2.49$, which is
gratifyingly close to the (frequentist) LS value $e_{LS}/\mu  = 2.45$ given
by Eq.(14) for $p_{e} = 0.05$. The thresholds for other 
critical levels are given in Table A.1, with the corresponding upper
limits (bounds) $\varepsilon_{U}$ in Table A.2. 

For the representative measurement $e/\mu = 1.0$ of Sect. 3.7,
the $95\%$ upper limit from Table A.2 is $2.41 \mu$. This exceeds the
corresponding values in Table 1, confirming that the uniform prior
is the more conservative option.

\begin{figure}
\vspace{8.2cm}
\includegraphics{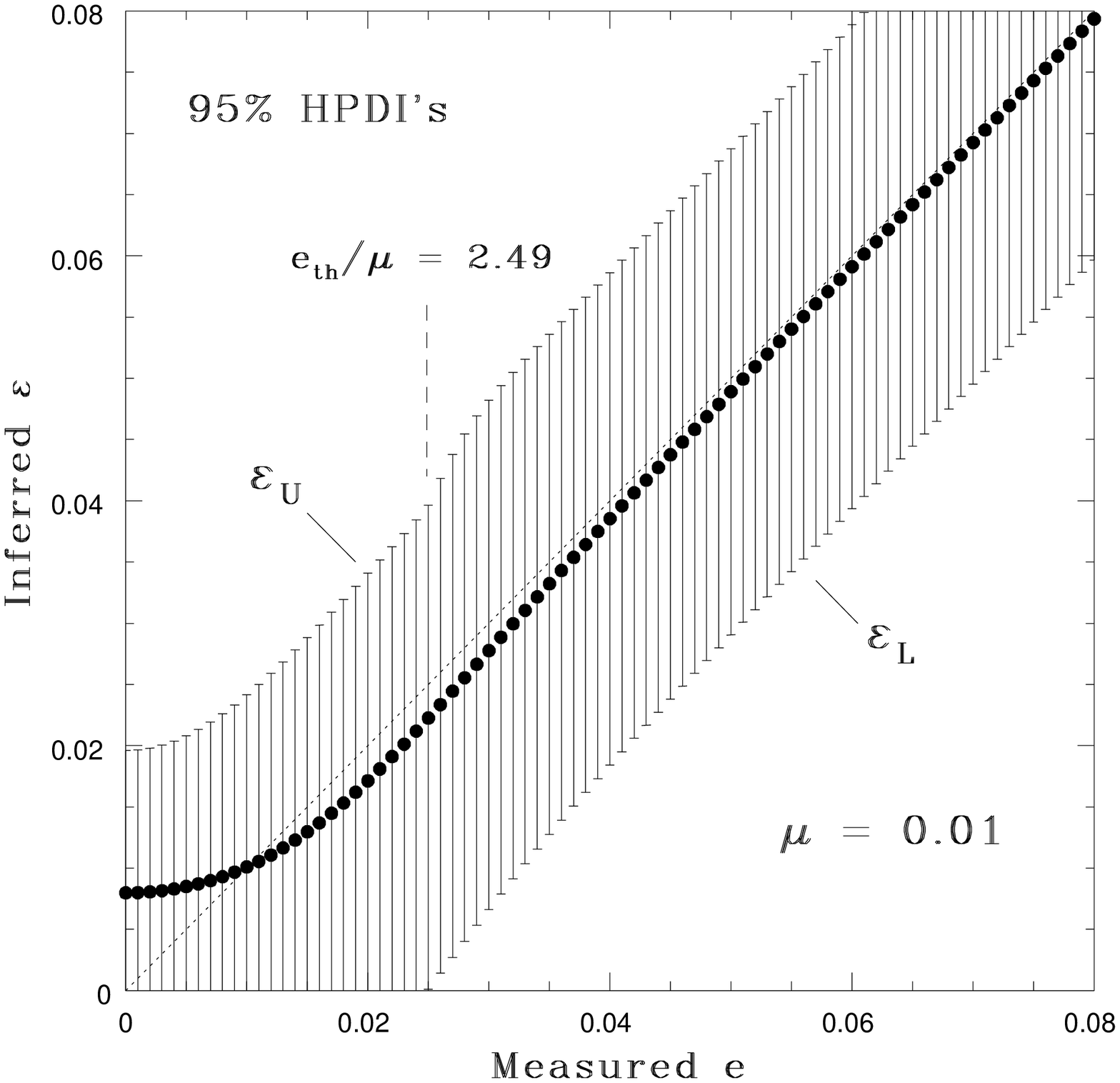}
\caption{Highest posterior density intervals for the posterior
pdf $q(\varepsilon|e)$. The enclosed posterior probability of each HPDI is 
$95\%$, and the prior pdf $\psi$ is given by Eq.(A.1). The filled circles are 
the posterior means $<\varepsilon>$ computed from Eq.(10).}
\end{figure}
\begin{figure}
\vspace{8.2cm}
\includegraphics{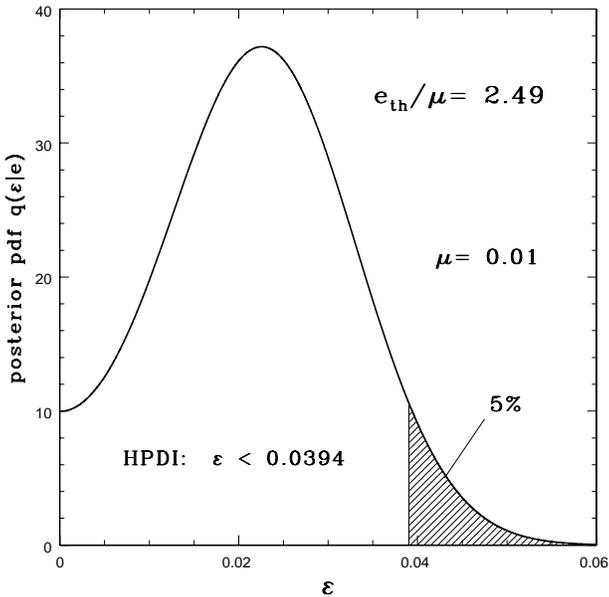}
\caption{Detection threshold. The posterior pdf $q(\varepsilon|e)$
for $e_{th} = 0.0249$, the measured value for marginal detection (Sect.3.5).
The hatched area with $\varepsilon > \varepsilon_{U} = 0.0394$ contains $5\%$
of the probability.
Note that $q(0|e_{th}) = q(\varepsilon_{U}|e_{th})$.}
\end{figure}

\begin{table}

\caption{Detection thresholds for $e$.}

\label{table:1}

\centering

\begin{tabular}{c c c }

\hline\hline

 $\alpha(\%)$   & $e_{LS}/\mu$   & $e_{th}/\mu$  \\

\hline
\hline

    31.7          & 1.52         &   1.71  \\

    10.0          & 2.15         &   2.21  \\

    \bf{5.0}      & \bf{2.45}    &   \bf{2.49}  \\

     1.0          & 3.03         &   3.06  \\

     0.1          & 3.72         &   3.74  \\

\hline
\hline

\end{tabular}

\end{table}

\begin{table}

\caption{Upper limits (bounds) for $\varepsilon$.}

\label{table:2}

\centering

\begin{tabular}{c c c c c c}

\hline\hline

 $e/\mu$  & $\varepsilon_{68.3}/\mu$ & $\varepsilon_{90}/\mu$  & 
 $\varepsilon_{\bf{95}}/\mu $    &  $\varepsilon_{99}/\mu$ & $\varepsilon_{99.9}/\mu$       \\

\hline
\hline

    0.0   & 1.00   &  1.64      & \bf{1.96}    &   2.58   &  3.29  \\

    0.5   & 1.06   &  1.75      & \bf{2.08}    &   2.73   &  3.48  \\

    1.0   & 1.28   &  2.05      & \bf{2.41}    &   3.11   &  3.89    \\

    1.5   & 1.66   &  2.50      & \bf{2.88}    &   3.59   &  4.38   \\

    2.0   & (2.59) &  3.03      & \bf{3.41}    &   4.12   &  4.90   \\

    2.5   & (3.31) &  (3.87)    & \bf{(3.96)}  &   4.65   &  5.43     \\

    3.0   & (3.86) &  (4.52)    & \bf{(4.82)}  &   5.18   &  5.96   \\

    3.5   & (4.37) &  (5.04)    & \bf{(5.36)}  &   (5.97) &  6.48     \\

    4.0   & (4.89) &  (5.54)    & \bf{(5.86)}  &   (6.49) & (7.18)    \\

\hline
\hline

\end{tabular}

\end{table}

\subsection{A revised Lucy-Sweeney test}

Even without evidence of additional components, 
an investigator may be reluctant to base an analysis of orbital elements
on uncertain estimates of tidal decay.  
If so, the assumption of a uniform prior for $\varepsilon$
should be attractive. 
Physically, this corresponds to no secular evolution of
$\varepsilon$ {\em and} a formation mechanism that uniformly populates the 
interval $0 < \varepsilon < 1$. Accordingly, a system with $e \approx 0$ is 
assumed to have formed as such (cf. Sect. 3.1).

This {\em neutral} standpoint is an attractive basis for a revised version of
the LS test in which the previous acceptance of a circular orbit ($H_{0}$) is
now replaced by an upper limit.

The revised LS test with $1-\alpha = 0.95$ proceeds as follows:\\

1) The eccentricity $e \pm \mu$ is derived from the least squares solution.\\

2) If $e/\mu > 2.49$, this measured value $e \pm \mu$ is accepted.\\

3) However, if $e/\mu < 2.49$, the measured value is replaced by the upper 
   limit $\varepsilon_{95}$ obtained by interpolation in column 4 of 
   Table A.2.\\

To illustrate this revised test, upper limits are given in Table A.3 for
six SB1's for which circular orbits are reported on the first page of Table 1
in LS. Thus for YZ Cas, the least squares value $e = 0.004$ was rejected by the
LS test and $\varepsilon = 0$ accepted. We now compute the $95\%$ upper limit 
as follows:
The estimate $\mu = 0.0037$ is derived from Eq.(13) and Table 1 of LS. 
Linear interpolation in Table A.2 at $e/\mu = 1.1$ then gives
$\varepsilon_{95}/\mu = 2.50$, whence 
$\varepsilon < \varepsilon_{95} = 0.009.$\\

Table A.3. shows that upper limits can differ by large factors, reinforcing
the earlier remark (Sect. 3.6) that upper limits are to be preferred in 
testing theories of tidal dissipation. A critical
data base of detections and upper limits would facilitate progress in this
field.

\begin{table}

\caption{Upper limits $\varepsilon_{95}$ for SB1 sample.}

\label{table:3}

\centering

\begin{tabular}{c c c c c c}

\hline\hline

 Star  & $e$ & $\mu$  & $e/\mu$ & $\varepsilon_{95}/\mu$ & $\varepsilon$ \\

\hline
\hline

    YZ Cas    & (0.004)   &  0.0037   & 1.10 & 2.50   &  $< 0.009$    \\

    HD 7345   & (0.046)   &  0.019   & 2.47 & 3.93   &  $< 0.073$    \\

    DM Per    & (0.09 )   &  0.074   & 1.22 & 2.62   &  $< 0.19$     \\

    HD 16589  & (0.008)   &  0.0087   & 0.92 & 2.35   &  $< 0.021$    \\

    HD 18337  & (0.073)   &  0.030   & 2.48 & 3.93   &  $< 0.12$     \\

    HD 21912  & (0.005)   &  0.011   & 0.45 & 2.07   &  $< 0.023$    \\

\hline
\hline

\end{tabular}

\end{table}

\acknowledgement

I am grateful to the referee for pointing out an error in statistical
terminology.

\end{document}